\documentclass[12pt]{article}

\usepackage{subfig} 
\usepackage{url}
\usepackage{wrapfig}
\usepackage{eufrak} 
\usepackage{amsmath}   
\usepackage{a4p}
\usepackage{microtype}\usepackage{lmodern}\usepackage[T1]{fontenc}
\usepackage{mathptmx}
\usepackage{graphicx}
\usepackage[normalem]{ulem} 
\usepackage{color}
\definecolor{myred}{rgb}{0.6,0,0} 
\definecolor{myblue}{rgb}{0,0.2,0.4}
\definecolor{mygreen}{rgb}{0,0.9,0.1}
\definecolor{Orange}{rgb}{1.,0.65,0.}
\usepackage{color}
\definecolor{myred}{rgb}{1.0,0,0} 
\definecolor{mygreen}{rgb}{0,0.9,0.1} 
\definecolor{myblue}{rgb}{0,0.2,0.4}
\definecolor{mygray}{rgb}{.8,.8,.8}
\definecolor{darkorange}{rgb}{1, 0.549, 0}
\definecolor{purple}{rgb}{0.6,0.4,0.6}
\definecolor{mymagenta}{rgb}{0.6,0.4,0.6}
 \definecolor{LightCyan} {rgb}{0.88,1.,1.}
 \definecolor{Orange} {rgb}{1.,0.65,0.}
 \definecolor{PaleGreen} {rgb}{0.6,0.98,0.6}
 \definecolor{Pink} {rgb}{1.,0.75,0.8}
\definecolor{Red}{rgb}{1,0,0}
   \definecolor{Blue}{rgb}{0,0,1}
   \definecolor{Yellow}{rgb}{1,1,0}
   \definecolor{Orange}{rgb}{1,0.4,0}
   \definecolor{Pink}{rgb}{1,0,1}
   \definecolor{Purple}{rgb}{0.5,0,0.5}
   \definecolor{Teal}{rgb}{0,0.5,0.5}
   \definecolor{Navy}{rgb}{0,0,0.5}
   \definecolor{Aqua}{rgb}{0,1,1}
   \definecolor{Lime}{rgb}{0,1,0}
   \definecolor{Green}{rgb}{0,0.5,0}
   \definecolor{Olive}{rgb}{0.5,0.5,0}
   \definecolor{Maroon}{rgb}{0.5,0,0}
   \definecolor{Brown}{rgb}{0.6,0.4,0.2}
   \definecolor{Black}{gray}{0}
   \definecolor{Gray}{gray}{0.5}
   \definecolor{Silver}{gray}{0.75}
   \definecolor{White}{gray}{1}

\usepackage{hyperref}  
\definecolor{darkblue}{rgb}{0,0,.5}
\hypersetup{
linktocpage=true,
    bookmarks=true,         
breaklinks=true,
    frenchlinks=false, %
    unicode=false,          
    pdftoolbar=true,        
    pdfmenubar=true,        
    pdffitwindow=false,     
    pdfstartview={FitH},    
    pdfauthor={J.Fleischer, T.Riemann},     
    pdfcreator={JF+TR},   
    pdfproducer={JF+TR}, 
    pdfkeywords={}, 
    pdfnewwindow=true,      
    colorlinks=false,       
    linkcolor=red,          
    citecolor=green,        
    filecolor=magenta,      
    urlcolor=cyan           
}
\usepackage[all]{hypcap} 

\usepackage{rotating}

\numberwithin{equation}{section}
\numberwithin{figure}{section}
\numberwithin{table}{section}

\newcommand{\be}{\begin{equation}}
\newcommand{\ee}{\end{equation}}
\newcommand{\bea}{\begin{eqnarray}}
\newcommand{\eea}{\end{eqnarray}}

\newcommand{\nl}{\nonumber \\}

\newcommand{\nn}{\nonumber}

\begin{document}
\sloppy

\texttt{
\begin{flushleft}
DESY 11-207
\\
BI-TP 2011/38  
\\
SFB/CPP-11-69
\\
LPN 11-65 
\\
\end{flushleft}
}
\vspace{1cm}

\bigskip

\begin{center}
{\LARGE \bf
A solution for tensor reduction of one-loop $N$-point functions with $N \ge 6$}
\\
\vspace{1.0cm}

\renewcommand{\thefootnote}{\fnsymbol{footnote}}
{\Large J. Fleischer~\footnote{E-mail:~fleischer@physik.uni-bielefeld.de}${}^{a}$  ~~and~~
        T. Riemann~\footnote{E-mail:~Tord.Riemann@desy.de}${}^{b}$        }
\\[1cm]
\end{center}

{\noindent
{${}^{a}$~Fakult\"at f\"ur Physik, Universit\"at Bielefeld, Universit\"atsstr. 25,  33615
Bielefeld, Germany }
\\\noindent
{${}^{b}$~Deutsches Elektronen-Synchrotron, DESY, Platanenallee
  6, 15738 Zeuthen, Germany}
}

\vspace{1cm}

\begin{center}
{\Large \bf
{Abstract}}
\\[5mm]\end{center}
Collisions at the LHC produce many-particle final states, and for precise predictions the one-loop $N$-point 
corrections are needed. We study here the tensor reduction for Feynman integrals with $N\ge 6$.
A general, recursive solution by Binoth et al. expresses  $N$-point Feynman integrals of rank $R$ 
in terms of $(N-1)$-point 
Feynman integrals of rank $(R-1)$ (for $N\ge 6$).
We show that the coefficients can be obtained analytically from suitable representations of the metric tensor.
Contractions of the tensor integrals with external momenta can be efficiently {expressed 
as well.
We consider our approach particularly well suited for automatization}.
\bigskip

PACS index categories: 12.15.Ji, 12.20.Ds, 12.38.Bx

\setcounter{footnote}{0}
\renewcommand{\thefootnote}{\arabic{footnote}}

\section{\label{Intro} Introduction}
In a recent article  \cite{Fleischer:2010sq}, we have worked out an algebraic method to present one-loop tensor 
$5$-point functions in terms of  scalar one-loop $1$-point to 
$4$-point functions.
The tensor integrals are defined as
\bea
\label{definition}
 I_{n}^{\mu_1\cdots\mu_R} &=&~\int \frac{d^dk}{i {\pi}^{d/2}}~~\frac{\prod_{r=1}^{R} k^{\mu_r}}{\prod_{j=1}^{n}c_j},
\eea
with denominators $c_j$, having  \emph{chords}
$q_j$,
\begin{eqnarray}\label{propagators}
c_j &=& (k-q_j)^2-m_j^2 +i \epsilon.
\end{eqnarray}
In a {subsequent} article \cite{Fleischer:2011nt} we have calculated contractions of the $N=5$-point tensor Feynman integrals {with external momenta},
resulting in the analytic evaluation of sums over products of
scalar products of the chords and signed minors, yielding 
compact expressions for them. 
{The present article is based on the observation that those sums are 
{valid for arbitrary $N$.}
This allows us to extend our formalism to $N$-point tensor integrals with $N\ge 6$.

Following ideas presented in \cite{Bern:1993kr}, an iterative approach has been systematically worked out in 
\cite{Binoth:2005ff}.
The  $N$-point tensor integrals are represented {there} in terms of 
 $(N-1)$-point tensor integrals with smaller rank for arbitrary $N \ge 6$ as
\bea
I_N^{{\mu}_1 {\mu}_2 \dots {\mu}_R } = -\sum_{r=1}^N C_r^{{\mu}_1} I_{N-1}^{{\mu}_2 \dots {\mu}_R ,r},
\label{recurs}
\eea
where $r$ indicates the line scratched from $I_N^{{\mu}_1 {\mu}_2 \dots {\mu}_R }$. 
Equation (61) of \cite{Binoth:2005ff} will be our starting point; it {contains an implicit solution for the coefficients $C_j^{\mu}$}:
\bea
\sum_{j=1}^N C_j^{\mu} q_j^{\nu} =\frac{1}{2} g_{[4]}^{{\mu\nu}} .
\label{Cs}
\eea
The subscript $[4]$, indicating explicitly the $4$-dimensional metric tensor, will be skipped in the following.
Further, we will set $q_N=0$ throughout this article, in notations of \cite{Binoth:2005ff} $r_N=0$,  and 
${\Delta}_{jN}^{\nu}=- q_j^{\nu} $. In the present work
we develop a procedure to solve \eqref{Cs} analytically for arbitrary $N\ge 6$. Explicit examples will be
given for $N=7$ and $N=8$.

Assume a representation of the metric tensor in the form 
\bea
\frac{1}{2} g^{\mu \nu}= \sum_{i,j=1}^{N-1} G_{ij} q_i^{\mu} q_j^{\nu}
\label{mett}
\eea
is available.
Then, necessarily, the vector
\bea
\label{Cj}
C_j^{\mu}=\sum_{i=1}^{N-1} G_{ij} q_i^{\mu} 
\eea
is a solution of \eqref{Cs}. An additional requirement according to eq. (62) in \cite{Binoth:2005ff} has to
be fulfilled for this vector:
\bea
\sum_{j=1}^N C_j^{\mu}=0,
\label{nanunana}
\eea
which we will verify for the solutions we obtain.

{Our approach consists in finding 
an object which, contracted
with chords $q_a$ and $q_b$, results in ($q_a \cdot q_b$). This yields {a} representation of the metric tensor
requested in \eqref{mett}, from which {the coefficients $C_j^{\mu}$ can be obtained}}. For an
$N$-point function with $N$ external momenta
one has to do with $(N-1)$ vectors, $q_N=0$. 
For $N=5$ one has in general 4 independent vectors from which one can construct uniquely
the metric tensor in $4$ dimensions. For $N > 5$ one has $(N-5)$ ``scratched'' vectors which are 
not supposed to enter the construction of the metric tensor. Nevertheless the contraction of the 
metric tensor with any pair of the available $(N-1)$ vectors must result in their scalar product. 
This is the problem we solve in the present article analytically. 

{It might be interesting here 
to recall the corresponding relation} for $5$-point functions 
\cite{Diakonidis:2009fx}, which can be considered as {an} inhomogeneous {analogue} to \eqref{recurs}:
\bea\label{3.52}
I_5^{{\mu}_1 \dots {\mu}_r}=
I_5^{{\mu}_1 \dots {\mu}_{r-1}} \cdot Q_0^{{\mu}_r}
-\sum_{s=1}^{5} I_4^{{\mu}_1 \dots {\mu}_{r-1}, s} \cdot Q_s^{{\mu}_r},
\eea
reducing $5$-point functions to $4$-point functions {of lower rank}.
Here the vectors \footnote{Throughout this article, greek indices are running from $0$ to $N$, while
latin indices are running from $1$ to $N$.}
\bea
\label{Q6}
Q_{\sigma}^{\mu}&=&\sum_{i=1}^{5}  q_i^{\mu} \frac{{{\sigma}\choose i}_5}{\left(  \right)_5},~~~ {\sigma}=0 \cdots 5, \\
\label{Q7}
Q_s^{t,\mu}&=&\sum_{i=1}^{5} q_i^{\mu} \frac{ {st\choose it}_5}{{t\choose t}_5},~~~ s,t=1 \cdots 5.
\eea
have been introduced and $\left( \dots \right)_5$ are Cayley determinants with elements 
\bea
Y_{ij}=-(q_i-q_j)^2+m_i^2+m_j^2.
\label{Yij}
\eea
In the following, setting up sums over scalar products multiplied with Cayley determinants,
we will implicitly use the relation 
\bea\label{Scalar2a}
(q_i \cdot q_j) =\frac{1}{2} \left[Y_{ij}-Y_{in}-Y_{nj}+Y_{nn} \right] ,
\eea
which is valid if $q_N=0$. Therefore we have to assume this 
from the very beginning.
\section{\label{tensor6}$6$-point functions}
$6$-point functions of arbitrary tensor rank are
well known, see e.g. \cite{Denner:2005nn,Diakonidis:2008ij}
\footnote{In \cite{Denner:2005nn} it is also indicated how $7$-point functions may be evaluated.}. Eq. \eqref{recurs} reads in this case
\bea
I_6^{{\mu}_1 {\mu}_2 \dots {\mu}_R}= -\sum_{r=1}^6 C_r^{{\mu}_1} I_5^{{\mu}_2 \dots {\mu}_R,r}.
\label{6point}
\eea 
In (4.6) of \cite{Diakonidis:2008ij}, the $C_r^{{\mu}} \equiv C_r^{s,{\mu}} = -v_r^{\mu}$ is given:
\bea
C_r^{{\sigma},{\mu}}=\sum_{i=1}^5 q_i^{\mu} \frac{{0r\choose {\sigma}i}_6}{{0\choose s}_6},~~~~~~~~{\sigma}=0 \dots 6.
\label{Qsr}
\eea
Here the index ${\sigma}$ indicates a certain redundancy, i.e. the vector $C_r$ {is not unique}.
This reflects the property of eq. (58) of \cite{Binoth:2005ff} to have only a ``pseudo-inverse''.
We now {will} verify this equation in order to demonstrate our approach. Due to \eqref{recurs}, we 
{have} to find a solution of \eqref{Cs}.
To be systematic, we first {collect} the following  set of sums, with $s=1 \dots n$:
\bea
\label{sum1}
\sum_{i,j=1}^{n-1} (q_a \cdot q_i) (q_b \cdot q_j) {0i\choose 0j}_n~&&=~\frac{1}{2}(q_a \cdot q_b){0\choose 0}_n
-\frac{1}{4}{\left( \right)}_n \left(Y_{an}-Y_{nn}\right) \left(Y_{bn}-Y_{nn}\right) ,
\\
\label{sum2}
\sum_{i,j=1}^{n-1} (q_a \cdot q_i) (q_b \cdot q_j) {0i\choose sj}_n~&&=~\frac{1}{2}(q_a \cdot q_b){0\choose s}_n
+\frac{1}{4}{\left( \right)}_n \left({\delta}_{as}-{\delta}_{ns}\right) \left(Y_{bn}-Y_{nn}\right) ,
\\
\label{sum3}
\sum_{i,j=1}^{n-1} (q_a \cdot q_i) (q_b \cdot q_j) {si\choose sj}_n~&&=~\frac{1}{2}(q_a \cdot q_b){s\choose s}_n
-\frac{1}{4}{\left( \right)}_n \left({\delta}_{as}-{\delta}_{ns}\right) \left({\delta}_{bs}-{\delta}_{ns}\right)
.
\eea
In fact \eqref{sum2} and \eqref{sum3} have been written in \cite{Fleischer:2011nt} for $n=5$, but 
as mentioned above it turns out that all
the sums written in \cite{Fleischer:2011nt} are as well valid for any $n$.
For $n=6$ {it is} ${\left( \right)}_6=0$.
Indeed equations \eqref{sum1}-\eqref{sum3} have to be considered as {identities} in terms of the $Y_{ij}$,
and the property ${\left( \right)}_6=0$ has to be taken into account as special property for the $Y_{ij}$ in  \eqref{Yij}.
Thus we can finally write the metric tensor as
\bea
\label{gum1}
\frac{1}{2} g^{\mu \nu} &=\sum_{i,j=1}^{5} \frac{1}{{0\choose 0}_6} {0i\choose 0j}_6 q_i^{\mu} q_j^{\nu} ,
\\
\label{gum2}
\frac{1}{2} g^{\mu \nu} &=\sum_{i,j=1}^{5} \frac{1}{{0\choose s}_6} {0i\choose sj}_6 q_i^{\mu} q_j^{\nu} ,
\\
\label{gum3}
\frac{1}{2} g^{\mu \nu} &=\sum_{i,j=1}^{5} \frac{1}{{s\choose s}_6} {si\choose sj}_6 q_i^{\mu} q_j^{\nu}
,
\eea
and due to \eqref{Cj}, equations \eqref{gum1} and \eqref{gum2} yield {the solution} \eqref{Qsr} for ${\sigma}=0$ and 
${\sigma=s}\ne0$, respectively,
while \eqref{gum3} yields another option; {see also (75) in \cite{Fleischer:1999hq}}.
{In this case} we have
\bea
C_r^{s,{\mu}}=\sum_{i=1}^5 q_i^{\mu} \frac{{sr\choose si}_6}{{s\choose s}_6}=Q_r^{s,\mu},~~~~~~~~s=1 \dots 6,
\label{Css}
\eea
in the notation of \eqref{Q7}.
We remark that ${s\choose s}_6$ in \eqref{Css} is a Gram determinant of a $5$-point function, which
 may become small in certain domains of phase space. Here, we have a certain choice, $s=1, \dots 6$, so that one presumably 
{will find} an $s$ for which ${s\choose s}_6$ is not small.

{
It is interesting to note that {the} $C_r$ in \eqref{Qsr} for $\sigma=0$ and $\sigma>0$ agree. 
{One may see this, starting} from the identity (see also (A.11) of \cite{Diakonidis:2008ij} or (A.13) of \cite{Melrose:1965kb})
\bea
\label{identity}
{0\choose s}_6 {0i\choose 0j}_6={0\choose 0}_6 {0j\choose si}_6+{0\choose i}_6 {0j\choose 0s}_6 
\eea
or
\bea
\frac{{0i\choose 0j}_6}{{0\choose 0}_6}=\frac{{0j\choose si}_6}{{0\choose s}_6}+{0\choose i}_6 \frac{{0j\choose 0s}_6}
{{0\choose 0}_6 {0\choose s}_6}.
\eea
Multiplying with $q_i^{\mu}$ and summing over $i$ proves our statement since
\bea
\label{ZeroSum}
\sum_{i=1}^5 q_i^{\mu} {0\choose i}_6 =0.
\eea
{The last identity} follows from the vanishing of all scalar products of \eqref{ZeroSum} with any non-vanishing chord;
see also (A.2) of \cite{Fleischer:2011nt}:
\bea
\sum_{i=1}^5 (q_a \cdot q_i) {0\choose i}_6 = -\frac{1}{2} (Y_{a6}-Y_{66}) \cdot \left( \right)_6 =0,~~~a=1, \dots 5.
\eea
A similar calculation shows that \eqref{Css} indeed differs from \eqref{Qsr}. The reason why different
representations are of interest is a possible optimization of the numerics: Some representations may have small
determinants in the denominator, while others don't.
}

{
It remains to verify \eqref{nanunana}. For \eqref{Qsr} with $\sigma=0$ we use
\bea
\sum_{r=1}^6 {0r\choose 0i}_6=-{0\choose i}_6
\eea
such that
\bea
\sum_{r=1}^6 C_r^{0, \mu} = \sum_{r=1}^6 \sum_{i=1}^{5} q_i^{\mu} \frac{{0r\choose 0i}_6}{{0\choose 0}_6}=
-\frac{1}{{0\choose 0}_6} \sum_{i=1}^{5} q_i^{\mu} {0\choose i}_6 =0 ,
\eea 
due to \eqref{ZeroSum}. For \eqref{Css} the proof is simpler since all we need is
\bea
\sum_{r=1}^6 {sr\choose si}_6=0, ~~~s=1, \dots 6.
\eea
}

\section{\label{tensor7}$7$-point functions}
{For} the $7$-point functions we first investigate the corresponding representation of \eqref{Qsr} for
$\sigma=0$. Eq.~(A.9) of \cite{Fleischer:2011nt} can be written  for arbitrary $n$ and $s=1 \dots n$
as an identity in terms of the $Y_{ij}$:
\bea
\label{A.9}
\sum_{i,j=1}^{n-1} (q_a \cdot q_i) (q_b \cdot q_j) {0si\choose 0sj}_n
&=&\frac{1}{2}(q_a \cdot q_b){0s\choose 0s}_n 
 \\
&&
-\frac{1}{4}\left\{\left[{s\choose s}_n \left(Y_{an}-Y_{nn}\right)+{s\choose 0}_n \left({\delta}_{as}-{\delta}_{ns} \right)
\right]\left(Y_{bn}-Y_{nn} \right) \right. 
\nn \\ \nn
&&+\left. \left[{s\choose 0}_n \left(Y_{an}-Y_{nn}\right)+{0\choose 0}_n \left({\delta}_{as}-{\delta}_{ns} \right)
\right]\left({\delta}_{bs}-{\delta}_{ns} \right) \right\}.
\eea
{As was observed} for the $6$-point function {in the discussion of} \eqref{sum1} - \eqref{sum3}, the vanishing of certain determinants
simplifies the result considerably. 
Quite generally with dimension $4$ of the chords,
all $\left( \right)_n$, $n \ge 7$, have rank $6$, i.e. any (signed) minor of order $7$ vanishes \cite{Melrose:1965kb}.
{The} $\left( \right)_7$ is of order $8$ and thus {also the} ${0\choose 0}_7$,
${s\choose 0}_7$ and ${s\choose s}_7$ vanish\footnote{See also \cite{Fleischer:1999hq}.}, and therefore the whole
curly bracket in \eqref{A.9} vanishes with the result 
\bea
\sum_{i,j=1}^{6} (q_a \cdot q_i) (q_b \cdot q_j) {0si\choose 0sj}_7~=~
\frac{1}{2}(q_a \cdot q_b){0s\choose 0s}_7. 
\eea
In general, {it is} ${0s\choose 0s}_7 \ne 0$ and {one} can write
\bea
\frac{1}{2} g^{\mu \nu} =\sum_{i,j=1}^{6} \frac{1}{{0s\choose 0s}_7}{0si\choose 0sj}_7  q_i^{\mu} q_j^{\nu} ,
\eea
from which we read off
\bea
C_r^{s0,\mu}=\sum_{i=1}^{6} \frac{1}{{0s\choose 0s}_7} {0si\choose 0sr}_7 q_i^{\mu}.
\label{C70}
\eea
This is exactly the result as in \eqref{Qsr} ($\sigma=0$), only that now a line and a column of the
${0\choose 0}_7$ is scratched in addition - a very natural result.
{
Even more, we also find the correspondence of \eqref{Qsr} for $\sigma=s >0$ in the form
\bea
C_r^{st,\mu}=\sum_{i=1}^{6} \frac{1}{{0s\choose ts}_7} {0si\choose tsr}_7 q_i^{\mu}.
\label{C77}
\eea
{In order} to show that \eqref{C70} and \eqref{C77} are equal, we proceed as for the $6$-point
function, starting from an \emph{extensional}\footnote{See \cite{Melrose:1965kb} for extensionals.} of \eqref{identity},
\bea
\label{idendidy}
{0s\choose ts}_7 {0sj\choose 0si}_7={0s\choose 0s}_7 {0sj\choose tsi}_7+{0s\choose is}_7 {0sj\choose 0st}_7 
\eea
or
\bea
\frac{{0sj\choose 0si}_7}{{0s\choose 0s}_7}=\frac{{0sj\choose tsi}_7}{{0s\choose ts}_7}+{0s\choose is}_7 \frac{{0sj\choose 0st}_7}
{{0s\choose 0s}_7 {0s\choose ts}_7}.
\eea
Multiplying with $q_i^{\mu}$ and summing over $i$ proves the statement since
\bea
\label{ZeroTum}
\sum_{i=1}^6 q_i^{\mu} {0s\choose is}_7 =0 .
\eea
{Eq. \eqref{ZeroTum}}  follows again from the vanishing of all scalar products of \eqref{ZeroTum} with any non-vanishing chords
as given in (A.6) of \cite{Fleischer:2011nt}:
\bea 
\label{eq-wa-83b}
{\Sigma}^{2,s}_{a} &\equiv&
\sum_{i=1}^{n-1}(q_a \cdot q_i)  {0s\choose is}_n 
~=~
-\frac{1}{2} \left\{
{s\choose s}_n\left(Y_{an}-Y_{nn}\right) +
{s\choose 0}_n\left({\delta}_{as}-{\delta}_{ns}\right)
\right\}.
\eea
For $n=7$ the order of the determinants on the right-hand side of \eqref{eq-wa-83b} is 7,
but their rank is 6 and thus they all vanish}.

Similarly we proceed for the {analogue of} \eqref{Css}, {which was proven for} the $6$-point function.
We start from a relation like \eqref{A.9},
which was not directly needed in \cite{Fleischer:2011nt}, but occurred {there} in an intermediate step:
\bea
\label{not}
\sum_{i,j=1}^{n-1} (q_a \cdot q_i) (q_b \cdot q_j) {sti\choose stj}_n &=& \frac{1}{2}(q_a \cdot q_b){st\choose st}_n 
 \\
&&-\frac{1}{4}
\left\{\left[{s\choose s}_n \left({\delta}_{at}-{\delta}_{nt} \right)-{s\choose t}_n \left({\delta}_{as}-{\delta}_{ns} \right)
\right]\left({\delta}_{bt}-{\delta}_{nt} \right) \right. 
\nn \\ \nn
&&+\left. \left[{t\choose t}_n \left({\delta}_{as}-{\delta}_{ns} \right)-{s\choose t}_n \left({\delta}_{at}-{\delta}_{ns} \right)
\right]\left({\delta}_{bs}-{\delta}_{ns} \right) \right\},
\eea
{with} $s,t=1 \dots n$.
According to the above discussion, {it is}
${s\choose s}_7={t\choose t}_7={s\choose t}_7=0$,
such that again the curly bracket in \eqref{not} vanishes. Since ${st\choose st}_7 \ne 0$
in general - {it is} a $5$-point Gram determinant - we can write
\bea
\frac{1}{2} g^{\mu \nu} =\sum_{i,j=1}^{6} \frac{1}{{st\choose st}_7}{sti\choose stj}_7  q_i^{\mu} q_j^{\nu} ,
\eea
from which we read off
\bea
C_r^{st,\mu}=\sum_{i=1}^{6} \frac{1}{{st\choose st}_7} {sti\choose str}_7 q_i^{\mu}=Q_r^{st,\mu} 
\label{C71}
\eea
in the notation of \eqref{Q6} and \eqref{Q7}.
Again this result corresponds exactly to \eqref{Css}, only that another line and column are scratched
from the ${s\choose s}_7$. 
 The upper indices $s$ and $t$ of $C_r^{st,\mu}$ in \eqref{C71} indicate again the redundancy
in the determination of this vector. {They} can be given any values {within} $s,t=1 \cdots 7$.
{In fact this vector differs again
from \eqref{C70} and \eqref{C77}. Finally, similarly {to the case of the} $6$-point function one proves \eqref{nanunana} 
for $C_r^{s0,\mu}$.
}

\section{\label{tensor8}$8$-point functions}
For the $8$-point functions - and analogeously for $N > 8$ - the calculation
follows the same lines as for the $7$-point function. 
Again, at first we investigate the representation corresponding to \eqref{Qsr} for $\sigma=0$. 
{Eq.} (A.13) of \cite{Fleischer:2011nt} can be written for $s,t=1 \dots n$ as 
\bea
\label{eq-wa-90}
{\Sigma}^{4,st}_{ab} &\equiv&
\sum_{i,j=1}^{n-1} (q_a \cdot q_i) (q_b \cdot q_j) {0sti\choose 0stj}_n
 \nn\\
&=& \frac{1}{2}(q_a \cdot q_b) {0st\choose 0st}_n 
 \\
&&- \frac{1}{4}~
\left\{\left[ {st\choose st}_n\left(Y_{an}-Y_{nn} \right)~
+{st\choose s0}_n\left({\delta}_{at}-{\delta}_{nt} \right)~
+{ts\choose t0}_n\left({\delta}_{as}-{\delta}_{ns} \right) \right] \left(Y_{bn}-Y_{nn} \right) \right. 
\nn \\
&&\left. ~+
\left[ {s0\choose s0}_n\left({\delta}_{at}-{\delta}_{nt} \right)~
+{st\choose s0}_n\left(Y_{an}-Y_{nn}  \right)~
-{s0\choose t0}_n\left({\delta}_{as}-{\delta}_{ns} \right) \right] \left({\delta}_{bt}-{\delta}_{nt} \right) \right. 
\nn \\ \nn
&&\left. ~+
\left[ {t0\choose t0}_n\left({\delta}_{as}-{\delta}_{ns} \right)~
-{s0\choose t0}_n\left({\delta}_{at}-{\delta}_{nt} \right)~
+{ts\choose t0}_n\left(Y_{an}-Y_{nn} \right)
 \right] \left({\delta}_{bs}-{\delta}_{ns} \right) 
\right\}.
\eea
For $n=8$, the determinants in the curly bracket of \eqref{eq-wa-90} are of order $7$ while their rank is only
$6$. Therefore they all vanish and so does the curly bracket. Since ${0st\choose 0st}_8 \ne 0$,
we obtain
\bea
\frac{1}{2} g^{\mu \nu} =\sum_{i,j=1}^{7} \frac{1}{{0st\choose 0st}_8}{0sti\choose 0stj}_8  q_i^{\mu} q_j^{\nu} ,
\eea
from which we read off
\bea
C_r^{0st,\mu}=\sum_{i=1}^{7} \frac{1}{{0st\choose 0st}_8} {0sti\choose 0str}_8 q_i^{\mu}
~=~Q_r^{0st,\mu}.
\label{C80}
\eea
Similarly, we proceed in order to obtain the relation
corresponding to \eqref{Css}. Here we start from (A.14) of \cite{Fleischer:2011nt}
which we write for $s,t,u=1 \dots n$ as
\bea
\label{eq-wa-92}
{\Sigma}^{4,stu}_{ab} &\equiv&
\sum_{i,j=1}^{n-1} (q_a \cdot q_i) (q_b \cdot q_j) {stui\choose stuj}_n
\nn\\
&=& \frac{1}{2}(q_a \cdot q_b) {stu\choose stu}_n  
\\
&&- \frac{1}{4}~
\left\{~~\left[ {st\choose st}_n\left({\delta}_{au}-{\delta}_{nu} \right)~
-{st\choose su}_n\left({\delta}_{at}-{\delta}_{nt} \right)~
-{ts\choose tu}_n\left({\delta}_{as}-{\delta}_{ns} \right) \right] \left({\delta}_{bu}-{\delta}_{nu} \right) \right. 
\nn \\
&&\left. ~+
\left[ {su\choose su}_n\left({\delta}_{at}-{\delta}_{nt} \right)~
-{st\choose su}_n\left({\delta}_{au}-{\delta}_{nu} \right)~
-{su\choose tu}_n\left({\delta}_{as}-{\delta}_{ns} \right) \right] \left({\delta}_{bt}-{\delta}_{nt} \right) \right. 
\nn \\\nn 
&&\left. ~+
\left[ {tu\choose tu}_n\left({\delta}_{as}-{\delta}_{ns} \right)~
-{su\choose tu}_n\left({\delta}_{at}-{\delta}_{nt} \right)~
-{ts\choose tu}_n\left({\delta}_{au}-{\delta}_{nu} \right)
 \right] \left({\delta}_{bs}-{\delta}_{ns} \right) 
\right\}.
\eea
{Again}, the curly bracket in \eqref{eq-wa-92} vanishes.
${stu\choose stu}_8$ is also the Gram determinant of a $5$-point function,
obtained from the $8$-point function under consideration by scratching lines and columns $s,t$ and $u$.
{It} does not vanish in general. Thus we obtain
\bea
\frac{1}{2} g^{\mu \nu} =\sum_{i,j=1}^{7} \frac{1}{{stu\choose stu}_8}{stui\choose stuj}_8  q_i^{\mu} q_j^{\nu} ,
\eea 
from which we read off
\bea
C_r^{stu,\mu}=\sum_{i=1}^{7} \frac{1}{{stu\choose stu}_8} {stui\choose stur}_8 q_i^{\mu}~=~Q_r^{stu,\mu} ,
\label{C81}
\eea
in the notation of \eqref{Q6} and \eqref{Q7}.
Here again the upper indices $s,t$ and $u$ stand for the redundancy of the vector and can be freely chosen. 
\section{{Contractions of tensor integrals} with external momenta}
In \cite{Fleischer:2011nt} and \cite{Fleischer:2011xy} we have advertised the contraction of tensor {integrals}
with external momenta
for the calculation of Feynman diagrams. This led us to {a systematic study of} sums over products of
contracted chords and signed 
minors. These {sums} have found in the present work a generalisation {by exploiting} the fact that
{they} are valid not only for the specific value $n=5$ {as was assumed} for the $5$-point functions.
{Indeed, they are} correct for any $n$. 
In {the following, we demonstrate that due to this property we cannot only derive as above specific 
representations of the metric tensor, but also the contraction with external momenta can be performed
in the same way for $N$-point functions with $N \ge 6$, quite similarly as it was done for $5$-point functions.
Just for the purpose of demonstration we confine ourselves to tensors up to rank $3$ of the $7$-point
function.

For the vector {integral} \eqref{recurs} yields
\bea
I_7^{\mu}=-\sum_{r=1}^{7}  C_r^{\mu} I_6^r,
\eea
where $C_r^{\mu} \equiv C_r^{st,\mu}$ can be chosen, e.g., from \eqref{C71}, and {the} $I_6^r$ is the scalar $6$-point 
function obtained from the
scalar $7$-point function by scratching line $r$. For the contraction of {$C_r^\mu$} with a  {chord $q_a^\mu$} we need
the generalization of (A.15) of \cite{Fleischer:2011nt} {for $n=7$}:
\bea
\label{eq-wa-86}
(q_a \cdot C_r^{st}) ~\equiv~
  \frac{{\Sigma}^{1,rst}_{a}}{{st\choose st}_n} &=& \frac{1}{{st\choose st}_n}
\sum_{i=1}^{n-1}(q_a \cdot q_i)  {sti\choose str}_n 
\\\nn 
&=&\frac{1}{2{st\choose st}_n}\left[
 {st\choose st}_n\left({\delta}_{ar}-{\delta}_{nr} \right)~
-{st\choose sr}_n\left({\delta}_{at}-{\delta}_{nt} \right)~
-{ts\choose tr}_n\left({\delta}_{as}-{\delta}_{ns} \right)  
\right] ,
\eea
{with $s,t,r=1 \dots n$}.
{Eq.} {\eqref{eq-wa-86} has a surprising consequence. According to {the} construction in \cite{Binoth:2005ff},
the original tensor remains unchanged no matter how the vector $C_r^{\mu}$ is chosen, as long as conditions \eqref{mett}
and \eqref{nanunana} are fulfilled. Thus, contracting $C_r$ with {some} chord $q_a$, one still {may select}  the \emph{redundancy indices} $s,t$. The optimal choice is apparently $s,t \ne a,N$. In this case
only the first term in the square bracket of \eqref{eq-wa-86} remains and moreover ${st\choose st}_n$ cancels,
i.e. the redundancy disappears and formally we can write
\bea
(q_a \cdot C_r) = \frac{1}{2} \left( {\delta}_{ar}-{\delta}_{Nr} \right).
\eea
{As a result,  only $2$ terms remain in the sum  \eqref{recurs} after the contraction .}
Since $C_r$ carries the first
index ${\mu}_1$ of tensors of any rank this scalar product  {will occur in all applications}.

For the tensor of rank $2$, eq. \eqref{recurs} yields
\bea
I_7^{\mu \nu}=-\sum_{r=1}^{7}  C_r^{\mu} I_6^{\nu,r} =-\sum_{r=1}^{7}  C_r^{\mu} \left[ -\sum_{t=1}^{7} 
\sum_{j=1}^{6} q_j^{\nu} \frac{{rst\choose rsj}_7}{{rs\choose rs}_7} \cdot I_5^{rt} \right],
\label{tens2}
\eea
where the square bracket is the $6$-point vector according to \eqref{6point}. For $C_t^s$ 
we have taken for convenience the form resulting from \eqref{Css}. 
{Here again \eqref{eq-wa-86}
with $n=7$ can be used for the projection of the $6$-point vector. The only freedom we have now, however,
is the choice of $s$. Contracting with another vector $q_b$ the choice $s \ne r,b,N$ is optimal with the
result ($N=7$)
\bea
\label{appearance}
q_b \cdot \sum_{j=1}^{N-1} q_j^{\nu} \frac{{rst\choose rsj}_N}{{rs\choose rs}_N}=\frac{1}{2} \left( {\delta}_{bt}-{\delta}_{Nt} \right)
-\frac{1}{2} \frac{{rs\choose ts}_N}{{rs\choose rs}_N} \left( {\delta}_{br}-{\delta}_{Nr} \right).
\eea
{Only the sums over the basic functions $I_{5}^{rt}$ survive.}
The reason for the appearance of the second term in \eqref{appearance} is that for $r=b$ the $6$-point function
as a scratched $7$-point function is contracted with a vector, {which is} not among the vectors defining the $6$-point
function, and for $r=N$ all $6$-point vectors are nonvanishing. Thus there remains the redundancy index $s$.

Similarly we proceed for the tensor of degree $3$:
\bea
I_7^{\mu \nu \lambda}=-\sum_{r=1}^{7}  C_r^{\mu} I_6^{\nu \lambda,r} =-\sum_{r=1}^{7}  C_r^{\mu} 
\left[ -\sum_{t=1}^{7} \sum_{j=1}^{6} q_j^{\nu} \frac{{rst\choose rsj}_7}{{rs\choose rs}_7}  \cdot I_5^{\lambda,rt}\right].
\label{tens3}
\eea
The vector of the $5$-point function can be written as \cite{Fleischer:2010sq}
\bea
I_5^{\lambda,rt}=- \sum_{i=1}^6 q_i^{\mu} \sum_{u=1}^7 \frac{{0rtu\choose 0rti}_7}{{0rt\choose 0rt}_7} I_4^{rtu}. 
\label{V5}
\eea
According to \eqref{V5},
we see that the only new sum needed for the projection is\footnote{{See also (A.17) of \cite{Fleischer:2011nt}.}}
\bea
\label{eq-wa-87}
{\Sigma}^{2,stu}_{a} 
&\equiv&
\sum_{i=1}^{n-1}(q_a \cdot q_i)  {0stu\choose 0sti}_n
~=~
\frac{1}{2} 
 \Biggl\{{stu\choose st0}_n  \left(Y_{an}-Y_{nn} \right)
\nl && 
+~{0st\choose 0st}_n\left({\delta}_{au}-{\delta}_{nu} \right)
-{0st\choose 0su}_n\left({\delta}_{at}-{\delta}_{nt} \right)
-{0ts\choose 0tu}_n\left({\delta}_{as}-{\delta}_{ns} \right) 
\Biggr\},
\eea
and $s,t,u=1 \dots n$, with $n=7$ {here}.

{We would like to close this section with few general remarks.}
From the very beginning we work with the original tensors, i.e. we do not consider cancellations of scalar products appearing
in the numerator {against propagators in the denominaotors}, i.e. we do not intend to cancel ``reducible'' terms.
{As} a consquence, no shifts of integration momenta are needed. 

{Additionally,} no shifts of integration momenta {are needed} in the iterations from $N$-point to $(N-1)$-point functions. This is {nicely} seen from \eqref{tens3} and \eqref{V5}. Working only with the original Gram determinant 
$\left( \right)_N$ of the diagram under consideration, the ``scratches'' are simply done in the
$\left( \right)_N$. For this purpose it is important to recall that all our sums are also
valid for the ``last'' value of $s,t, \dots =N$.

Assume that in \eqref{V5} $r,t \ne N$, then four vectors contribute in the sum for the
$5$-point vector, i.e. $q_N=0$ is still {valid}.
{If} otherwise $r=N$, {then} the line $N$ is scratched and $q_N=0$ is not available anymore. In this case
$5$ vectors $q_i$ contribute in the sum for the vector of the $5$-point function, which is just what is needed 
here \cite{Fleischer:2010sq},
since the general integration rules of course allow all chords to be different from $0$.

{In order to take advantage of the above formalism, the contractions of tensor integrals with external momenta should be implemented in a software package as building blocks,  which are parameterised by the chords}, $a,b \dots = 1 \dots (N-1)$. 
In fact it does not matter if $4$ or $5$ vectors contribute since this is automatically taken into
account by the Kronecker $\delta$-functions in the sums like \eqref{eq-wa-86} and \eqref{eq-wa-87}.
{If a calculation is organized such that the tensor integrals are contracted exclusively by external momenta (or chords), a quite efficient numerics will result.} 

\section{\label{conclude}Conclusions}
For {tensor $N$-point functions} of rank $N \ge 6$ we found an analytic form for the coefficients $C_r^{{\mu}_1}$ {appearing in} the iterative scheme \eqref{recurs}. The crucial point in the derivation is
the observation that sums over products of scalar products of chords and signed minors, derived earlier for $n=5$,
also hold identically for arbitrary $n$. 
Applying {recurrence} relations  directly to the evaluation of higher rank tensors, the appearance
of vanishing Cayley determinants makes their application cumbersome, {as indicated in \cite{Fleischer:1999hq}}.
In {the} 
present approach the  vanishing of certain Cayley determinants is obviously quite welcome,
reducing the sums for the vectors $C_r^{\mu}$ considerably. As a result we obtain, for any $N$, expressions with
inverse $5$-point Gram determinants with an arbitrary choice of the scratched lines.
Also {the inverses}  ${0s \cdots\choose 0s \cdots}_N$ and  ${0t \cdots\choose st \cdots}_N$
are admitted.
The sums {at our disposal} also allow to perform {contractions} of the tensor integrals
with external momenta {quite efficiently}. 
In particular use can be made of the freedom to choose the vectors $C_r^{{\mu}}$ to find
the optimal form for these contractions.
For higher rank tensors this property appears of particular relevance
because to each tensor index corresponds a summation over the nonvanishing chords. {For} the tensor of rank $7$ of the $7$-point function, e.g., this corresponds to $6^7=279936$ terms. 
{All these terms are summed here} 
analytically to a product of $7$ terms, which will certainly save a big amount of computer time and
storage. 

\section*{Acknowledgements}
The authors are grateful to G. Heinrich {and S. Dittmaier} for communication. J.F. thanks DESY for kind hospitality.
Work is supported in part by Sonderforschungsbereich/Trans\-re\-gio SFB/TRR 9 of DFG
``Com\-pu\-ter\-ge\-st\"utz\-te Theoretische Teil\-chen\-phy\-sik" 
and European Initial Training Network LHCPHENOnet PITN-GA-2010-264564.

\small


\begin{thebibliography}{10}
\expandafter\ifx\csname url\endcsname\relax
  \def\url#1{\texttt{#1}}\fi
\expandafter\ifx\csname urlprefix\endcsname\relax\def\urlprefix{URL }\fi
\expandafter\ifx\csname href\endcsname\relax
  \def\href#1#2{#2} \def\path#1{#1}\fi

\bibitem{Fleischer:2010sq}
J.~Fleischer, T.~Riemann, {Complete algebraic reduction of one-loop tensor
  Feynman integrals}, Phys. Rev. D83 (2011) 073004.
\newblock \href {http://arxiv.org/abs/1009.4436} {\path{arXiv:1009.4436}},
  \href {http://dx.doi.org/10.1103/PhysRevD.83.073004}
  {\path{doi:10.1103/PhysRevD.83.073004}}.

\bibitem{Fleischer:2011nt}
J.~Fleischer, T.~Riemann, {Calculating contracted tensor Feynman integrals},
  Phys.Lett. B701 (2011) 646--653.
\newblock \href {http://arxiv.org/abs/1104.4067} {\path{arXiv:1104.4067}},
  \href {http://dx.doi.org/10.1016/j.physletb.2011.06.033}
  {\path{doi:10.1016/j.physletb.2011.06.033}}.

\bibitem{Bern:1993kr}
Z.~Bern, L.~J. Dixon, D.~A. Kosower, {Dimensionally regulated pentagon
  integrals}, Nucl. Phys. B412 (1994) 751--816.
\newblock \href {http://arxiv.org/abs/hep-ph/9306240}
  {\path{arXiv:hep-ph/9306240}}, \href
  {http://dx.doi.org/10.1016/0550-3213(94)90398-0}
  {\path{doi:10.1016/0550-3213(94)90398-0}}.

\bibitem{Binoth:2005ff}
T.~Binoth, J.~Guillet, G.~Heinrich, E.~Pilon, C.~Schubert, An algebraic /
  numerical formalism for one-loop multi-leg amplitudes, JHEP 10 (2005) 015.
\newblock \href {http://arxiv.org/abs/hep-ph/0504267}
  {\path{arXiv:hep-ph/0504267}}, \href
  {http://dx.doi.org/10.1088/1126-6708/2005/10/015}
  {\path{doi:10.1088/1126-6708/2005/10/015}}.

\bibitem{Diakonidis:2009fx}
T.~Diakonidis, J.~Fleischer, T.~Riemann, J.~B. Tausk, {A recursive reduction of
  tensor Feynman integrals}, Phys. Lett. B683 (2010) 69--74.
\newblock \href {http://arxiv.org/abs/0907.2115} {\path{arXiv:0907.2115}},
  \href {http://dx.doi.org/10.1016/j.physletb.2009.11.049}
  {\path{doi:10.1016/j.physletb.2009.11.049}}.

\bibitem{Denner:2005nn}
A.~Denner, S.~Dittmaier, Reduction schemes for one-loop tensor integrals, Nucl.
  Phys. B734 (2006) 62--115.
\newblock \href {http://arxiv.org/abs/hep-ph/0509141}
  {\path{arXiv:hep-ph/0509141}}, \href
  {http://dx.doi.org/10.1016/j.nuclphysb.2005.11.007}
  {\path{doi:10.1016/j.nuclphysb.2005.11.007}}.

\bibitem{Diakonidis:2008ij}
T.~Diakonidis, J.~Fleischer, J.~Gluza, K.~Kajda, T.~Riemann, J.~Tausk, {A
  complete reduction of one-loop tensor 5- and 6-point integrals}, Phys. Rev.
  D80 (2009) 036003.
\newblock \href {http://arxiv.org/abs/0812.2134} {\path{arXiv:0812.2134}},
  \href {http://dx.doi.org/10.1103/PhysRevD.80.036003}
  {\path{doi:10.1103/PhysRevD.80.036003}}.

\bibitem{Fleischer:1999hq}
J.~Fleischer, F.~Jegerlehner, O.~Tarasov, Algebraic reduction of one-loop
  {Feynman} graph amplitudes, Nucl. Phys. B566 (2000) 423--440.
\newblock \href {http://arxiv.org/abs/hep-ph/9907327}
  {\path{arXiv:hep-ph/9907327}}, \href
  {http://dx.doi.org/10.1016/S0550-3213(99)00678-1}
  {\path{doi:10.1016/S0550-3213(99)00678-1}}.

\bibitem{Melrose:1965kb}
D.~B. Melrose, Reduction of {Feynman} diagrams, Nuovo Cim. 40 (1965) 181--213.
\newblock \href {http://dx.doi.org/10.1007/BF028329}
  {\path{doi:10.1007/BF028329}}.

\bibitem{Fleischer:2011xy}
J.~Fleischer, T.~Riemann, {Simplifying 5-point tensor reduction}, Acta Phys.
  Polon. B42 (2011) 2371--2378.
\newblock \href {http://arxiv.org/abs/1111.4153} {\path{arXiv:1111.4153}},
  \href {http://dx.doi.org/10.5506/APhysPolB.42.2371}
  {\path{doi:10.5506/APhysPolB.42.2371}}.

\end{thebibliography}

 \end{document}